\documentclass[sigconf]{acmart}
\AtBeginDocument{%
  }
\pdfoutput=1
\copyrightyear{2025}
\acmYear{2025}
\setcopyright{acmlicensed}\acmConference[MM '25]{Proceedings of the 33rd ACM International Conference on Multimedia}{October 27--31, 2025}{Dublin, Ireland}
\acmBooktitle{Proceedings of the 33rd ACM International Conference on Multimedia (MM '25), October 27--31, 2025, Dublin, Ireland}
\acmDOI{10.1145/3746027.3755808}
\acmISBN{979-8-4007-2035-2/2025/10}
\usepackage{listings}
\usepackage{stfloats}
\usepackage{tikz}
\usetikzlibrary{arrows.meta, positioning}
\usepackage{algorithm}
\usepackage{algpseudocode}
\usepackage{tabularx}
\usepackage{xcolor}
\usepackage{balance}
\acmSubmissionID{6126}

\usepackage{graphicx}
\usepackage{multirow}
\usepackage{subcaption}
\usepackage{makecell}
\usepackage{seqsplit} 
\usepackage{xurl}      

\newboolean{showcomments}
\setboolean{showcomments}{true}

\ifthenelse{\boolean{showcomments}}
{ \newcommand{\mynote}[3]{
   \protect\fbox{\bfseries\sffamily\scriptsize#1}
   {\small$\blacktriangleright$\textsf{\emph{\color{#3}{#2}}}$\blacktriangleleft$}}}
{ \newcommand{\mynote}[3]{}}

\newcommand{\dirk}[1]{\mynote{Dirk}{#1}{blue}}

\renewcommand{\dirk}[1]{} 
\begin{document}

\title{INDS: Incremental Named Data Streaming for Real-Time Point Cloud Video}

\author{Ruonan Chai}
\affiliation{%
  \institution{The Hong Kong University of Science and Technology (Guangzhou)}
  \department{Information Hub}
  \city{Guangzhou}
  \country{China}}
\email{rchai327@connect.hkust-gz.edu.cn}

\author{Yixiang Zhu}
\affiliation{%
  \institution{The Hong Kong University of Science and Technology (Guangzhou)}
  \department{Information Hub}
  \city{Guangzhou}
  \country{China}}
\email{g1312136480@163.com}

\author{Xinjiao Li}
\affiliation{%
  \institution{The Hong Kong University of Science and Technology (Guangzhou)}
  \department{Information Hub}
  \city{Guangzhou}
  \country{China}}
\email{xli886@connect.hkust-gz.edu.cn}

\author{Jiawei Li}
\affiliation{%
  \institution{The Hong Kong University of Science and Technology (Guangzhou)}
  \department{Information Hub}
  \city{Guangzhou}
  \country{China}}
\email{jli526@connect.hkust-gz.edu.cn}

\author{Zili Meng}
\affiliation{%
  \institution{The Hong Kong University of Science and Technology}
  \department{Department of Electronic and Computer Engineering}
  \city{Hong Kong}
  \country{China}}
\email{zilim@ust.hk}

\author{Dirk Kutscher}
\affiliation{%
  \institution{The Hong Kong University of Science and Technology (Guangzhou)}
  \department{Information Hub}
  \city{Guangzhou}
  \country{China}}
\email{dku@hkust-gz.edu.cn}

\begin{abstract}


Real-time streaming of point cloud video, characterized by massive data volumes and high sensitivity to packet loss, remains a key challenge for immersive applications under dynamic network conditions. While connection-oriented protocols such as TCP and more modern alternatives like QUIC alleviate some transport-layer inefficiencies, including head-of-line blocking, they still retain a coarse-grained, segment-based delivery model and a centralized control loop that limit fine-grained adaptation and effective caching. We introduce INDS (Incremental Named Data Streaming), an adaptive streaming framework based on Information-Centric Networking (ICN) that rethinks delivery for hierarchical, layered media. INDS leverages the Octree structure of point cloud video and expressive content naming to support progressive, partial retrieval of enhancement layers based on consumer bandwidth and decoding capability. By combining time-windows with Group-of-Frames (GoF), INDS’s naming scheme supports fine-grained in-network caching and facilitates efficient multi-user data reuse. INDS can be deployed as an overlay, remaining compatible with QUIC-based transport infrastructure as well as future Media-over-QUIC (MoQ) architectures, without requiring changes to underlying IP networks. Our prototype implementation shows up to 80\% lower delay, 15–50\% higher throughput, and 20–30\% increased cache hit rates compared to state-of-the-art DASH-style systems. Together, these results establish INDS as a scalable, cache-friendly solution for real-time point cloud streaming under variable and lossy conditions, while its compatibility with MoQ overlays further positions it as a practical, forward-compatible architecture for emerging immersive media systems.

\end{abstract}

\ccsdesc[500]{Networks~Network architecture design}
\ccsdesc[300]{Networks~Content delivery networks}
\ccsdesc[100]{Computer systems organization~Embedded and cyber-physical systems}
\ccsdesc[100]{Applied computing~Multimedia streaming}

\keywords{Point Cloud Streaming, Information-Centric Networking (ICN), Incremental Transmission, Adaptive Video Delivery}



\maketitle

\section{Introduction}
In recent years, immersive volumetric video has gained significant traction in applications such as remote collaboration, immersive communication, and extended reality (XR), demonstrating substantial market potential\cite{zhang2024networked}. Industry forecasts project that the global volumetric video market will reach USD 7.6 billion by 2028\cite{hu2025livevv}. However, 3D point cloud video presents considerable challenges for transmission, storage, and computation due to its massive data volume and highly dynamic nature. Compared to conventional 2D video, point cloud video can reach bitrates up to 2.9 Gbps and features high interactivity and rapid viewpoint switching, making efficient delivery over bandwidth-limited and lossy networks a pressing research challenge\cite{lee2020groot}. 

To manage the data complexity, point clouds are commonly compressed using Octree-based encoding~\cite{schnabel2006octree}. This method recursively subdivides 3D space and represents point cloud data as a tree structure, where parent nodes define spatial occupancy and leaf nodes store detailed voxel attributes. Due to the hierarchical dependency, decoding any subtree requires the successful reception of its parent nodes. However, this hierarchical structure puts forward new requirements for the transmission mechanism, that is, how to flexibly select and gradually obtain the required layer data according to the network status. 

Modern protocols like QUIC address TCP’s transport inefficiencies~\cite{iyengar2021quic, bishop2022rfc}, but DASH-style systems still rely on connection-oriented, segment-based delivery with fixed-bitrate representations and centralized control loops. For point cloud video, with its deeply hierarchical structure and unbalanced data distribution, such coarse-grained adaptation results in redundant transmission, poor cache utilization, and reduced resilience under dynamic or lossy conditions. Moreover, their reliance on a single-server, single-path paradigm further restricts scalability and hinders multi-user content distribution, especially in heterogeneous environments requiring rapid viewpoint adaptation.

Building on this foundation, DASH, originally designed for 2D video~\cite{sodagar2011mpeg, stockhammer2011dynamic}, has been extended to point cloud systems such as DASH-PC~\cite{hosseini2018dynamic}, PCC-DASH~\cite{van2019towards}, GROOT~\cite{lee2020groot}, and ViVo~\cite{han2020vivo}. These approaches improve adaptive bitrate switching and CDN caching, yet they remain constrained by fixed-bitrate ladders that hinder fine-grained adaptation and limit cache efficiency, leaving scalability challenges unresolved in multi-user scenarios.

While many multimedia applications like DASH are inherently information-centric, streaming coarse-grained objects over traditional host-centric networks limits this paradigm’s potential~\cite{6231276}. In contrast, native Information-Centric Networking (ICN) offers secure access to named content and enables application-layer control over transport functions such as reliability and flow control — features particularly attractive for media delivery. By leveraging multi-destination forwarding, in-network caching, and adaptive Interest-based retrieval, ICN enhances bandwidth efficiency and reduces end-to-end latency, especially under dynamic or lossy conditions~\cite{jacobson2009networking, ghasemiInternetScaleVideoStreaming2021, hanAMVSNDNAdaptiveMobile2013}. Importantly, these benefits can be achieved incrementally, through overlay deployment similar to CDNs and recent Media-over-QUIC~\cite{ietf-moq-overview}, without requiring fundamental changes to today’s IP infrastructure.  

Building on this foundation, prior studies such as AMVS-NDN~\cite{hanAMVSNDNAdaptiveMobile2013}, adaptive video streaming with network coding~\cite{wuAdaptiveVideoStreaming2020,saltarinAdaptiveVideoStreaming2017,leiNDNIoTContent2018}, and Internet-scale video streaming over NDN~\cite{ghasemiInternetScaleVideoStreaming2021} have shown that ICN can reduce server load, enable multi-destination and multi-source retrieval, and improve resilience to packet loss by avoiding end-to-end retransmissions. However, these works have primarily targeted 2D video, leaving the challenges of 3D point cloud streaming, including layered encoding, incremental transmission, and cache-aware delivery — largely unexplored.  

To address these gaps, this paper proposes INDS (Incremental Named Data Streaming), a dynamic layered point cloud streaming mechanism based on ICN. INDS improves bandwidth efficiency, cache reuse, and transmission stability by storing only the highest-resolution version at the producer and dynamically delivering the required incremental data upon client requests. Compared with TCP/IP-based approaches such as DASH-PC~\cite{hosseini2018dynamic} and PCC-DASH~\cite{van2019towards}, our method leverages ICN’s multi-destination transmission, intelligent caching, and adaptive layered delivery to significantly enhance overall performance. Moreover, INDS’s design is conceptually aligned with emerging Media-over-QUIC (MoQ) architectures: its chunk-level Interest expressions map to MoQ’s object-based transport, its in-network caching corresponds to reusable MoQ media objects, and its multi-destination forwarding mirrors MoQ’s publish/subscribe model. Although a full MoQ-based prototype is beyond the scope of this work, these synergies suggest that INDS can be incrementally deployed on top of MoQ without requiring changes to the underlying IP infrastructure.

The main contributions of this work are as follows:
\begin{itemize}
    \item We propose INDS, a novel ICN-based framework for real-time point cloud streaming. Leveraging a hierarchically structured naming scheme with in-network caching, INDS enables cache-assisted multi-destination delivery in lossy networks, reducing reliance on costly end-to-end retransmissions.

    \item We design a time-window and Group-of-Frames (GoF) based naming mechanism that enables fine-grained data organization, minimizes cache entry overhead, and improves both cache efficiency and scalability in multi-user environments.
    
    \item We adopt a single-version storage model combined with smart sampling, enabling dynamic adjustment of data granularity based on user bandwidth conditions and decoding capabilities, thereby reducing redundancy and improving transmission efficiency.
\end{itemize}

The remainder of this paper is organized as follows. Section~\ref{related} reviews related work on point cloud video delivery over TCP/IP and ICN. Section~\ref{octree} presents Octree-based Point Cloud organization. Section~\ref{inds} presents the proposed INDS mechanism, including hierarchical Interest queries, producer-side incremental responses, and cache reconstruction strategies. Section~\ref{section5} demonstrates experimental evaluation. Finally, Section~\ref{section6} concludes the paper and discusses future directions.

\section{RelatedWork}\label{related}

\subsection{Point Cloud Streaming over TCP/IP}

Traditional point cloud video streaming primarily relies on the TCP/IP protocol, leveraging MPEG-DASH as the foundation for adaptive bitrate control. DASH-PC~\cite{hosseini2018dynamic} extends the DASH architecture to support point cloud streaming over HTTP/TCP. It provides compatibility with existing internet infrastructure and manages multiple quality versions using Media Presentation Description (MPD) files. However, its caching mechanism only supports complete frames, lacking support for fine-grained point cloud caching. Furthermore, due to its single-server architecture, DASH-PC suffers from server overload under high user demand and lacks dynamic encoding adaptation, resulting in inefficient bandwidth utilization.

PCC-DASH~\cite{van2019towards} applies DASH to 6DoF point cloud streaming by integrating MPEG V-PCC compression and adaptive bitrate selection. It introduces three bitrate control strategies---Greedy, Uniform, and Hybrid---based on user position, viewing direction, and buffer status. While these strategies improve QoE, PCC-DASH's foundation on TCP and coarse-grained segment fetching prevents fine-grained, layer-selective retrieval. This fundamentally limits its adaptation flexibility and caching efficiency, as it still requires transmitting pre-defined, monolithic quality versions.

GROOT~\cite{lee2020groot} improves volumetric video streaming through a PD-Tree structure and GPU-based parallel decoding, achieving real-time playback at 30 FPS on mobile devices. Techniques such as JPEG compression and frustum-based visibility filtering reduce data transmission and improve rendering quality. However, GROOT still relies on HTTP/TCP, which introduces latency and scalability issues. It pre-encodes multiple quality versions and cannot adjust encoding dynamically, leading to redundant transfers and low cache efficiency.

ViVo~\cite{han2020vivo} proposes a visibility-aware point cloud streaming framework that reduces 41.8\%--70.1\% of transmission data while maintaining high visual quality (SSIM > 0.99). It combines viewport prediction and k-d tree compression (Draco) to reduce bandwidth requirements in practical networks such as 5G, LTE, and WiFi. Nonetheless, ViVo inherits TCP's single-path limitations and CDN-only caching, lacking fine-grained adaptivity or content-aware redundancy elimination.

In a similar direction, a rate-utility optimized volumetric streaming system~\cite{park2019rate} introduces 3D tiling, window-based buffering, and bandwidth-aware rate allocation for AR/VR scenarios. The system dynamically adjusts point cloud quality based on user viewport, occlusion, and distance, reducing unnecessary data transfer. However, it also operates over TCP/IP and traditional CDN, suffering from blocking under congestion and limited distributed caching capabilities.

While QUIC improves over TCP in terms of transport responsiveness, existing streaming systems such as DASH-PC and PCC-DASH still operate at the level of predefined representations and entire segment retrievals. They lack support for partial layer selection or in-network retransmission suppression, which are critical for bandwidth-efficient and loss-resilient point cloud streaming.

\subsection{ICN-based Video Streaming for 2D and IoT Applications}
ICN is based on the notion of {\em accessing named data in the network}, 
with Named Data Networking (NDN) being its most prominent and widely implemented variant~\cite{zhang2014named}. While our system is designed in the context of ICN in general, most prior research in this space—particularly for video delivery—has adopted NDN-specific semantics and protocols. For consistency with the literature, we refer to these systems using their original NDN terminology, but our analysis applies more broadly to ICN-based architectures.

NDN has been explored as a promising alternative to TCP/IP for video delivery due to its content-centric model, in-network caching, and multi-destination support. AMVS-NDN~\cite{hanAMVSNDNAdaptiveMobile2013} integrates DASH with NDN for adaptive mobile video streaming. It employs hybrid bitrate adaptation strategies based on throughput, buffer state, and network prediction, while enabling device-to-device cache sharing to reduce cellular load.

Saltarin et al.~\cite{saltarinAdaptiveVideoStreaming2017} propose a network coding-enabled NDN framework that improves bandwidth usage and cache hit ratio by encoding data at intermediate nodes. This architecture enhances robustness in lossy environments and reduces redundant transmissions. However, it is tailored for 2D video and does not consider the layered or hierarchical structure required for efficient point cloud delivery.

Lei et al.~\cite{leiNDNIoTContent2018} design an NDN content distribution strategy for 5G IoT, combining RLNC (Random Linear Network Coding) with probabilistic multi-destination forwarding (Pinform). The strategy improves bandwidth efficiency and cache utilization by 10--20\%, and boosts throughput by 50--100\%. Although not specific to video, its encoding and forwarding mechanisms provide valuable insights for scalable point cloud data distribution.

Rainer et al.~\cite{rainerInvestigatingPerformancePullBased2016} analyze pull-based adaptive streaming in NDN, evaluating different forwarding and caching policies (e.g., SAF, iNRR, CEE, PC). Their findings show that NDN outperforms TCP/IP in throughput and cache efficiency, though unstable load balancing may cause frequent bitrate switching. Their observation that larger cache sizes improve playback stability is particularly relevant to point cloud applications.

Wu et al.~\cite{wuAdaptiveVideoStreaming2020} propose a dynamic NDN multicast-based adaptive video streaming solution (NM-ABR) optimized for WLAN. By dynamically adjusting multicast rates and combining it with SVC (Scalable Video Coding), the scheme improves throughput and reduces stalling. This layered approach offers potential applicability to point cloud data segmented by quality or resolution.

Finally, Ghasemi et al.~\cite{ghasemiInternetScaleVideoStreaming2021} evaluate Internet-scale video streaming over a global NDN testbed. Their work demonstrates that in-network caching and multi-destination delivery reduce server load and improve playback quality. However, they note that conventional ABR logic requires adaptation to better fit NDN's data retrieval model.

While promising, existing ICN systems largely target 2D video or IoT data.
There remains a gap in research targeting large-scale, hierarchical 3D point cloud streaming over NDN, which our work aims to address.
However, none of these ICN-based approaches address the combined challenges of layered, hierarchical point cloud structures, adaptive enhancement selection, and cache-aware incremental retrieval.

\section{Octree-Based Point Cloud Organization}
\label{octree}

\subsection{Point Cloud Data Structure and Octree Organization}

A 3D point cloud is a set of spatial data points. Its massive data volume, reaching up to 3.6~Gbps uncompressed at 30~FPS, presents significant transmission and storage challenges~\cite{clemm2020toward}. To manage this complexity, systems widely adopt \textbf{Octree encoding}~\cite{schnabel2006octree}, a hierarchical compression method that recursively subdivides 3D space.

While Octree offers high compression efficiency, as shown in Table~\ref{tab:octree_compression}, and supports multi-resolution access~\cite{d20178i}, its structure introduces critical dependencies: decoding any part of the point cloud requires the successful reception of all its parent nodes in the tree. Furthermore, the data distribution is highly unbalanced. As detailed in Table~\ref{tab:color_compression}, leaf nodes can contain between 50\% and 90\% of the total data, while the top-layer nodes with essential structural information remain small. This structure demands a transmission strategy optimized for both data dependency and unbalanced distribution.

\begin{table}[ht]
\centering
\renewcommand{\arraystretch}{1.2}
\caption{Octree Compression Efficiency on Four Point Cloud Datasets}
\label{tab:octree_compression}
\begin{tabular}{cccc}
\toprule
\textbf{Dataset} & 
\makecell[c]{\textbf{Raw Size}\\\textbf{(MB)}} & 
\makecell[c]{\textbf{Encoded Size}\\\textbf{(MB)}} & 
\textbf{Compression Ratio} \\
\midrule
Longdress   & 82.68  & 2.07  & 97.50$\times$ \\
Soldier     & 720.62 & 1.87  & 99.74$\times$ \\
Loot        & 16.22  & 0.58  & 96.44$\times$ \\
Redandblack & 14.61  & 0.60  & 95.92$\times$ \\
\bottomrule
\end{tabular}
\end{table}
\vspace{-3pt}

\begin{table}[ht]
\centering
\renewcommand{\arraystretch}{1.1}
\begin{tabular}{ccc c}
\toprule
\textbf{Dataset} & \textbf{Coloring Tech.} & \textbf{Color Size} & \textbf{Last Level Size} \\
\midrule
\multirow{3}{*}{Longdress} 
  & JPEG &  48.05\%    & 51.76\% \\
  & AVIF &  23.00\%    & 76.71\%                          \\
  & WebP &  44.76\%    & 60.12\%                           \\
\midrule
\multirow{3}{*}{Soldier} 
  & JPEG &  49.60\%    & 50.11\%          \\
  & AVIF &  18.35\%    & 81.18\%                            \\
  & WebP &  37.41\%    & 62.22\%                           \\
\midrule
\multirow{3}{*}{Loot} 
  & JPEG &   28.86\%   & 70.51\%          \\
  & AVIF &   8.18\%    & 91.02\%                           \\
  & WebP &   18.53\%   &  80.75\%                          \\
\midrule
\multirow{3}{*}{Redandblack} 
  & JPEG &   34.43\%   & 64.97\%          \\
  & AVIF &   13.34\%   & 85.87\%                           \\
  & WebP &   30.21\%   & 72.97\%                           \\
\bottomrule
\end{tabular}
\captionsetup{skip=15pt}
\caption{Comparison of color compression ratios and last-layer sizes across datasets}
\label{tab:color_compression}
\vspace{-2em}
\end{table}

\subsection{Transmission Challenges and Problem Analysis}

In traditional TCP/IP-based systems such as those using the DASH architecture for point cloud streaming, clients typically request complete frame segments based on predefined representations. Each representation corresponds to a fixed combination of encoding layers, and clients must retrieve the entire version rather than selectively requesting partial layers based on current network conditions. This coarse-grained, segment-level transmission leads to redundant data retrieval, especially under constrained bandwidth or when only low-fidelity rendering is needed, resulting in inefficient bandwidth utilization, an approach that is particularly ill-suited for hierarchically structured data like Octree point clouds.

Moreover, since DASH operates over HTTP/TCP, reliability is ensured by TCP’s end-to-end retransmission and in-order delivery mechanisms. In the event of packet loss, TCP typically retransmits only the missing segments of the byte stream, rather than the entire application-layer object. However, these retransmissions, combined with congestion control mechanisms that reduce the sending window, significantly increase segment retrieval latency and may cause playback stalls. This behavior is particularly problematic for large octree-compressed point cloud segments, where even partial packet loss can delay the timely reconstruction of frames. To accommodate heterogeneous network conditions, existing DASH systems often prestore multiple full-quality versions for client switching. However, this multi-version strategy imposes significant storage overhead on the server side and limits cache reuse, while preventing clients from issuing fine-grained, segment-level adaptive requests.

Additionally, the Octree’s leaf nodes account for the majority of the total data volume. Although this layer has higher redundancy tolerance, naive transmission strategies—such as con-secutive segment loss—can cause noticeable degradation in local geometry or color fidelity. Current methods lack fine-grained management and intelligent sampling strategies for this layer, which limits display quality in bandwidth-limited environments. 

To address these challenges and fully exploit the transmission potential of Octree encoding, we propose a dynamic, incremental transmission mechanism based on ICN in the next chapter. 

\section{Design and Implementation of INDS}
\label{inds}

\subsection{Motivation and System Overview}

As thoroughly analyzed in Section 3.2, traditional streaming architectures are ill-suited for hierarchical point cloud data due to challenges like coarse-grained segment delivery and retransmission-induced latency. To overcome these limitations, we propose INDS (Incremental Named Data Streaming), an ICN-based framework tailored for octree-encoded point cloud delivery.

NDS does not require re-encoding or version management at the producer and can operate over standard ICN forwarders, making it compatible with current testbeds and overlay deployment models.

Furthermore, TCP’s end-to-end retransmission mechanism results in significant latency under high packet loss. Packet loss triggers TCP's retransmission mechanism, which, combined with congestion control, can significantly delay the reception of the complete segment, thereby degrading responsiveness. INDS allows users to incrementally request only the necessary data layers based on real-time network bandwidth and rendering capability.

\begin{figure}[ht]
    \centering
    \includegraphics[width=0.99\linewidth]{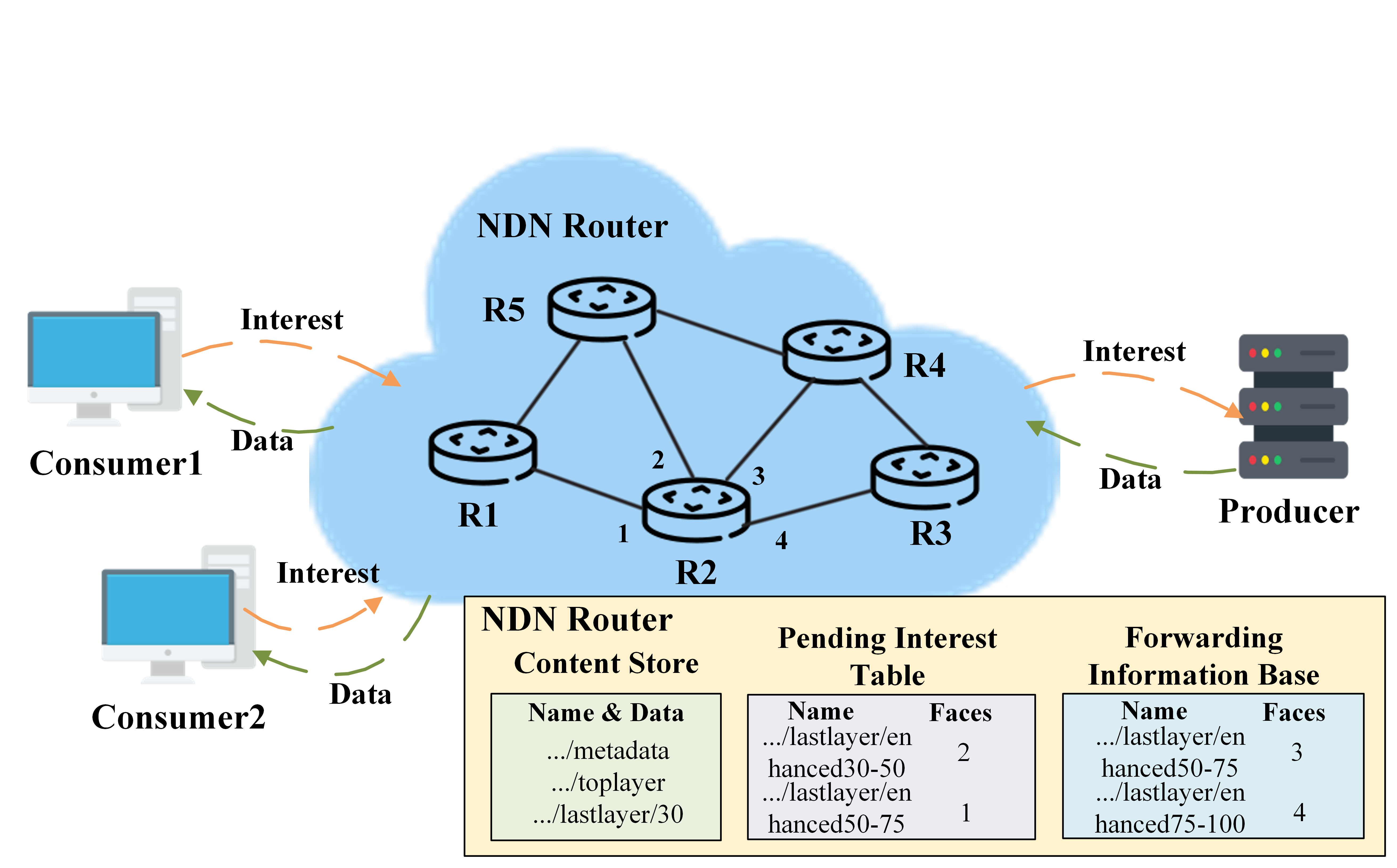}
    \caption{System Architecture of INDS.}
    \label{fig:inds_architecture}
\end{figure}

The overall system structure of INDS is depicted in Fig.~\ref{fig:inds_architecture}, consisting of consumers, NDN routers, and a producer. Consumers generate Interest packets adaptively. NDN routers forward, cache, and aggregate Interests using Content Store (CS), Pending Interest Table (PIT), and Forwarding Information Base (FIB).

The producer only stores the highest-quality content and responds passively to Interests without re-encoding. Consumers start by requesting a base segment (e.g., \texttt{/30}), and progressively request \texttt{enhanced30-50}, \texttt{enhanced50-75}, etc., when conditions improve. This progressive retrieval aligns with ICN's strength in chunk-level transmission.

\subsection{Robust Sampling and Hierarchical Naming}

To improve robustness under conditions of limited bandwidth and packet loss, we apply uniform sampling to the \texttt{LastLayer}. 
Specifically, we set three retention ratios of 30\%, 50\%, and 75\% 
and uniformly downsampled the original \texttt{LastLayer}, with 100\% retained as the full baseline for comparative analysis of reconstruction quality under different sampling rates. This sampling preserves the overall structural continuity of the point cloud, thereby avoiding concentrated local visual loss.

\begin{figure}[ht]
    \centering
    \includegraphics[width=0.48\textwidth]{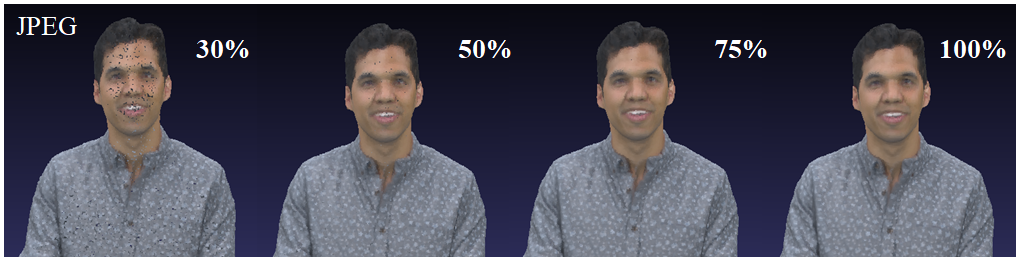}
    \includegraphics[width=0.48\textwidth]{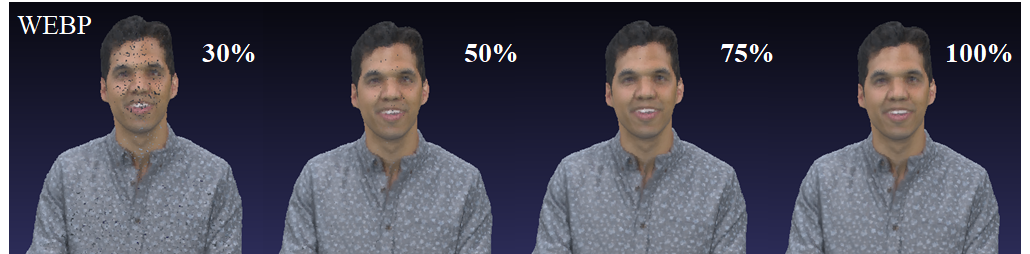}
    \includegraphics[width=0.48\textwidth]{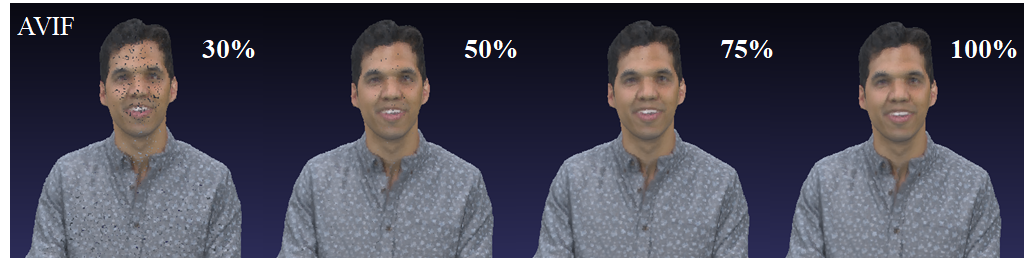}
    \caption{Reconstructed point cloud under different sampling rates using JPEG (top), WebP (middle), and AVIF (bottom).}
    \label{fig:recon_quality}
\end{figure}

From the experimental results shown in Fig.~\ref{fig:recon_quality}, even if only 30\% of the data is retained, the overall perceptibility of the geometric structure can be maintained. As the retention rate increases, the boundaries and color details are significantly restored. In addition, the uniform sampling strategy makes the residual data points evenly distributed in space, reducing the risk of severe image degradation caused by local packet loss. This sampling ensures graceful degradation and enables bandwidth-aware scheduling. Unlike typical segment-based approaches, spatially uniform sampling avoids concentrated loss artifacts.

To support incremental retrieval, INDS adopts the hierarchical naming scheme illustrated in Fig.~\ref{fig:ndn_naming}, which incorporates temporal windows and Group-of-Frames (GoF) to organize content at fine granularity. For example, segments such as \texttt{/LastLayer/30} and \texttt{enhanced30-50} represent progressive reconstruction levels. These examples are illustrative; INDS applies the same processing logic consistently across all enhancement segments.

\begin{figure}[ht]
    \centering
    \tiny  
    \renewcommand{\arraystretch}{1.5} 
    \setlength{\tabcolsep}{3pt} 
    \begin{tabular}{|p{1.0\linewidth}|}
        \hline
        /PointCloudService/DataSetID/MetaData \\ \hline
        /PointCloudService/DataSetID/TimeWindow\_20240314T120000/GoF\_0001/TopLayer \\ \hline
        /PointCloudService/DataSetID/TimeWindow\_20240314T120000/GoF\_0001/LastLayer/30 \\ \hline
        /PointCloudService/DataSetID/TimeWindow\_20240314T120000/GoF\_0001/LastLayer/enhanced30-50 \\ \hline
        /PointCloudService/DataSetID/TimeWindow\_20240314T120000/GoF\_0001/LastLayer/enhanced50-75 \\ \hline
        /PointCloudService/DataSetID/TimeWindow\_20240314T120000/GoF\_0001/LastLayer/enhanced75-100 \\ \hline
    \end{tabular}
    \caption{Hierarchical Naming for Incremental Point Cloud Delivery.}
    \label{fig:ndn_naming}
\end{figure}

\begin{figure*}[t]
    \centering
    \includegraphics[width=0.95\textwidth]{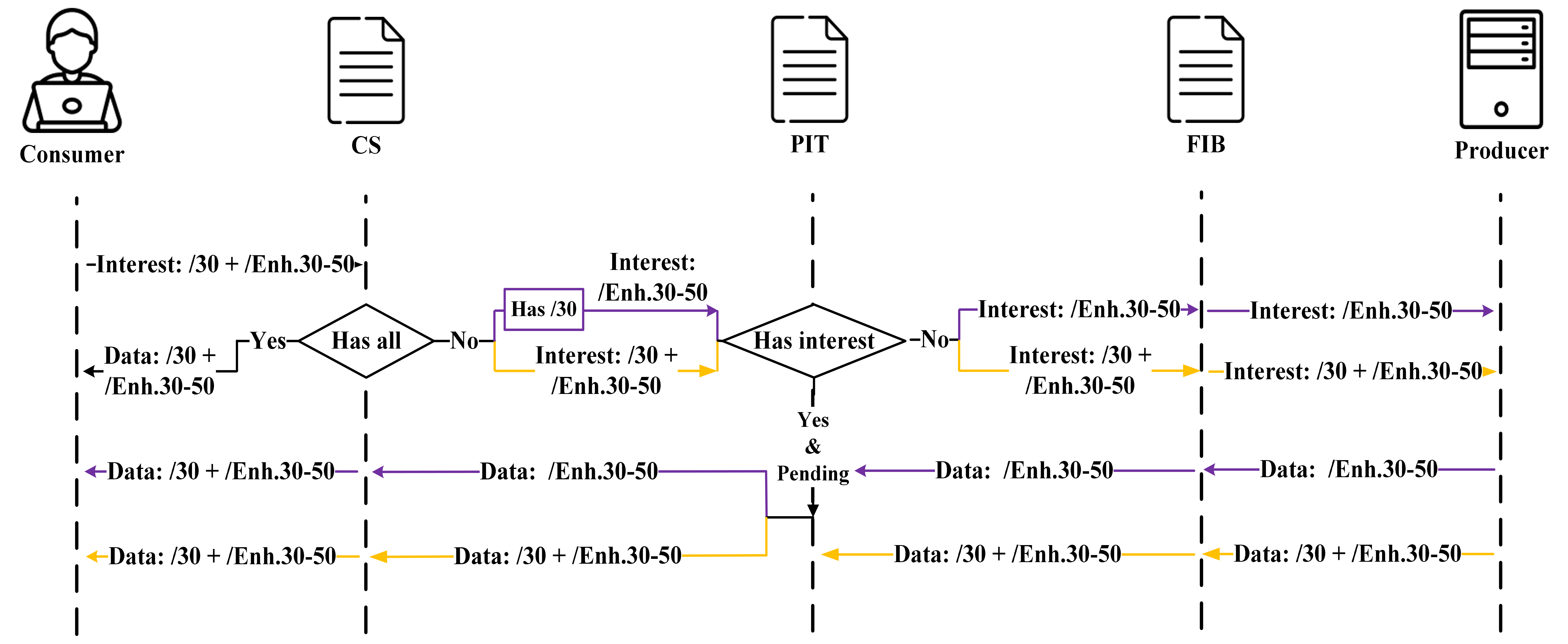}
    \caption{Progressive Interest/Data Exchange in INDS.}
    \label{fig:ndn_flow}
\end{figure*}

To further clarify the operational process, the complete Interest-Data workflow is illustrated in Fig.~\ref{fig:ndn_flow}, highlighting how adaptive Interests drive progressive transmission and how ICN routers efficiently utilize cached data to reduce load and latency. To illustrate the end-to-end Interest/Data workflow in INDS clearly, The process follows the algorithmic workflow described in Algorithm~\ref{alg:inds_protocol}.

\begin{algorithm}[ht]
\caption{INDS Interest/Data Exchange Protocol}
\label{alg:inds_protocol}
\begin{algorithmic}[1]
\Require Requested segments $S = \{\texttt{/30}, \texttt{/enh.30-50}\}$
\Ensure Return matching Data or forward Interest upstream

\Statex \textbf{Consumer Node:}
\State Construct Interest packet $I = S$clearly
\State Send $I$ to local Content Store (CS)

\Statex \textbf{Content Store (CS):}
\If{CS.has(\texttt{/30}) \textbf{and} CS.has(\texttt{/enh.30-50})}
    \State \Return Data(\texttt{/30}, \texttt{/enh.30-50}) \Comment{Full cache hit}
\ElsIf{CS.has(\texttt{/30})}
    \State \Return Data(\texttt{/30}) \Comment{Partial hit}
    \State Forward Interest(\texttt{/enh.30-50}) to PIT
\Else
    \State Forward Interest($I$) to PIT \Comment{Full miss}
\EndIf

\Statex \textbf{Pending Interest Table (PIT):}
\If{PIT.has($I$)}
    \State PIT.aggregate($I$) \Comment{Suppress duplicate Interests}
\Else
    \State PIT.insert($I$)
    \State Forward to FIB
\EndIf

\Statex \textbf{Forwarding Information Base (FIB):}
\State Select next hop(s) based on name prefix
\State Forward $I$ toward producer

\Statex \textbf{Producer Node (on receiving Interest $I$):}
\State Lookup requested segments $S$
\State Prepare corresponding Data packets
\State Send Data via reverse path

\end{algorithmic}
\end{algorithm}

\section{Results and Evaluation}\label{section5}
\label{evaluation}
\subsection{Testbed and Dataset}
\dirk{"realistic conditions" may not the best way to describe emulation-based environments. You could better say "To validate our design and to analyze the actual protocol behavior, we implemented a fully-functional prototype of INDS and tested it in an emulated network (using the MiniNDN platform).}
To validate our design and to analyze the actual protocol behavior, we implemented a fully-functional prototype of INDS and tested it in an emulated network using the MiniNDN platform\cite{marian2017minindn}. Mini-NDN employs Linux containers to run actual NFD forwarders, ensuring that forwarding, naming, and in-network caching behaviors are faithfully reproduced. Only the underlying hardware is virtualized, allowing experiments to capture realistic ICN protocol behavior rather than simplified simulation. Experiments are conducted with compressed point cloud data in simulated network scenarios.

The source data is taken from the public \textit{8i Voxelized Full Bodies (8iVFB v2)} dataset~\cite{d20178i}, a benchmark commonly used for evaluating point cloud compression and streaming. Four representative dynamic sequences are selected: \textit{Longdress}, \textit{Loot}, \textit{RedandBlack}, and \textit{Soldier}. Each sequence spans 10 seconds at 30 frames per second (fps), yielding 300 frames.

We focus on 1-second segments with 30 consecutive frames to evaluate per-second transmission behavior. For instance, the \textit{Loot} sequence compresses 484 MB of raw data to 8.61 MB using Octree-based encoding, resulting in an average compressed frame size of approximately 290 KB. The Octree depth is fixed at 7, allowing progressive transmission and level-of-detail control.

The ICN testbed includes one producer, ten consumers, and ten intermediate forwarding nodes (fwd0–fwd9). Nodes run NFD with in-network caching enabled. Bandwidths of 10, 50, and 80 Mbps are configured, with packet loss rates from 0\% to 1\% used to test robustness. All consumers continuously request point cloud frames at 30 fps. To ensure fair comparison, parallel TCP/IP-based topologies are constructed. For DASH-PC and PCC-DASH, a three-tier CDN architecture is built in Mininet\cite{lantz2010network}, comprising one core switch, three aggregation switches, and three access switches. One content source and three CDN caches reside at the aggregation/access layers. Ten consumers are distributed across access switches with matched bandwidth profiles. All link delays and capacities are aligned with the ICN setup for parity.

\subsection{Reconstruction Quality}

To quantify the reconstruction quality of INDS under progressive transmission, we evaluate two standard metrics—Peak Signal-to-Noise Ratio (PSNR) and Structural Similarity Index (SSIM)—across four enhancement levels: \texttt{/30}, \texttt{/50}, \texttt{/75}, and \texttt{/100}. Each level cumulatively includes more data segments, e.g., \texttt{/50} = \texttt{/30} + \texttt{enhanced30-50}.

\begin{figure}[ht]

    \centering
    \begin{subfigure}[b]{0.49\linewidth}
        \centering
        \includegraphics[width=\linewidth]{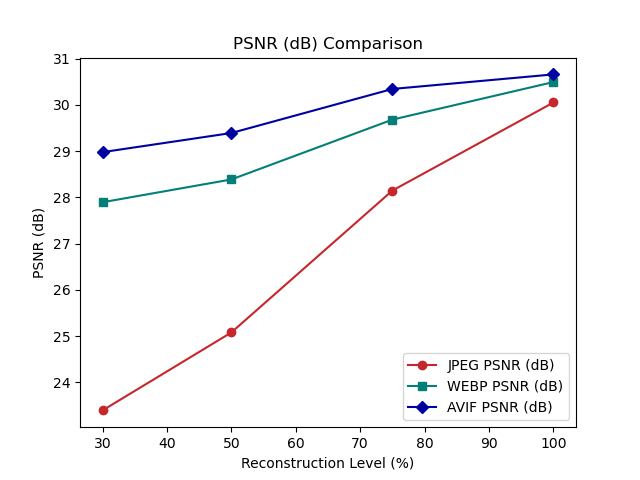}
        \caption{PSNR across four enhancement levels and three compression formats.}
        \label{fig:psnr_plot}
    \end{subfigure}
    \hfill
    \begin{subfigure}[b]{0.49\linewidth}
        \centering
        \includegraphics[width=\linewidth]{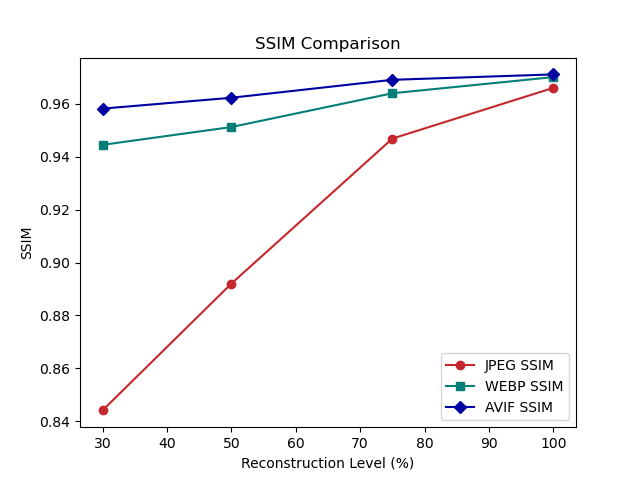}
        \caption{SSIM across four enhancement levels and three compression formats.}
        \label{fig:ssim_plot}
    \end{subfigure}
    
    \caption{Reconstruction quality evaluation using (a) PSNR and (b) SSIM under four enhancement levels (/30, /50, /75, /100) and three compression formats.}
    \label{fig:recon_quality_metrics}
\end{figure}

 Fig.~\ref{fig:recon_quality_metrics} reports the PSNR and SSIM scores for three common color compression formats—JPEG, WebP, and AVIF—under each enhancement level. As expected, both metrics exhibit steady improvements with increasing reconstruction level, demonstrating that INDS achieves fine-grained and controllable quality refinement. For example, JPEG’s PSNR increases from 23.5~dB at \texttt{/30} to over 30~dB at \texttt{/100}, while AVIF reaches 29.0~dB at the lowest level and peaks near 30.8~dB. SSIM shows similar growth, with AVIF consistently surpassing other formats at each stage.

Among all formats, AVIF outperforms JPEG and WebP across both PSNR and SSIM, particularly at lower enhancement levels. This advantage stems from AVIF's superior chroma subsampling and modern compression algorithms (e.g., AV1 intra prediction), which retain more color detail and structural integrity under bandwidth-constrained transmission.

\subsection{Bandwidth Adaptability Evaluation}
\label{sec:bandwidth-adapt}

We evaluate INDS's adaptability to varying bandwidth conditions by simulating three downstream rates: 10, 50, and 80 Mbps. The consumer continuously streams at 30~fps and dynamically selects enhancement segments based on runtime bandwidth estimates. For each frame, the consumer logs the requested segment types and the number of data packets successfully retrieved across different layers.

\begin{figure}[ht]
    \centering
    \includegraphics[width=0.5\textwidth]{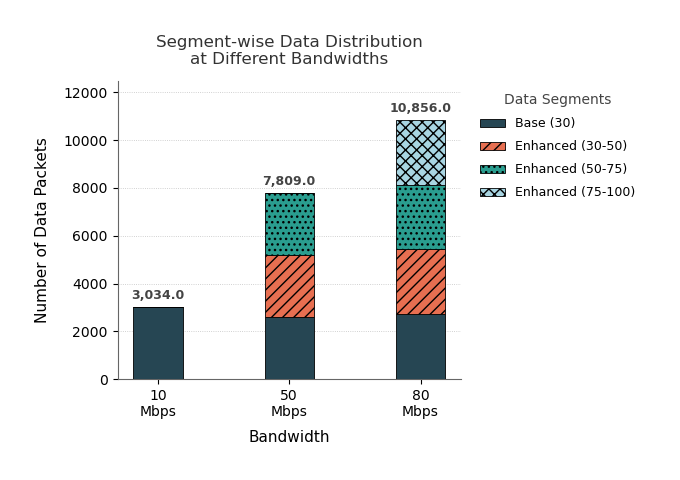}
    \caption{Segment-wise data packet distribution at 10, 50, and 80~Mbps.}
    \label{fig:adaptive}
\end{figure}

As shown in Fig.~\ref{fig:adaptive}, the total number of data packets increases with available bandwidth: from 3{,}034 packets (only \texttt{/30} base segment) at 10 Mbps, to 7{,}809 packets (including \texttt{enhanced30-50} and \texttt{enhanced50-75}) at 50 Mbps, and finally to 10{,}856 packets at 80 Mbps with all four segments retrieved. This progressive trend confirms that INDS gracefully scales its transmission granularity according to bandwidth availability.

Unlike DASH-style approaches that switch between coarse-grained pre-encoded versions, INDS supports fine-grained, segment-level adaptation through hierarchical naming and Interest-based requests. This results in more efficient bandwidth utilization, reduced redundancy, and improved tolerance to bandwidth fluctuations. Furthermore, INDS eliminates the need to store multiple versions at the producer, improving both scalability and caching efficiency in multi-consumer scenarios.
This demonstrates INDS’s ability to offer finer adaptation steps than traditional DASH-style switching, resulting in smoother quality scaling and better bandwidth utilization.

\subsection{Cache Efficiency Evaluation}
\label{sec:cache-efficiency}

We evaluate the cache efficiency of INDS against DASH-PC and PCC-DASH under packet loss rates from 0\% to 1\%. INDS achieves high cache efficiency primarily due to its fine-grained, semantically structured naming scheme, which enables each chunk to be reused across different users and sessions within the same temporal window. Caches employ a standard LRU policy (65,536 packets per node), consistent with NFD’s default configuration. This design allows INDS to perform segment-wise transmission for selective reuse, whereas TCP/IP-based systems cache only at frame-level granularity, limiting effectiveness under dynamic conditions.

As a result, INDS achieves a cache hit rate consistently between 73\% and 76\%, even as the packet loss rate increases to 1\%. Its time-window and Group-of-Frames (GoF) based naming scheme enables multiple users to reuse cached content from the same temporal segment. Moreover, consumers dynamically request only the necessary enhancement segments, such as \texttt{enhanced30-50}, based on real-time network conditions, avoiding redundant data retrieval and increasing content diversity in caches.

\begin{figure}[ht]
    \centering
    \includegraphics[width=0.5\textwidth]{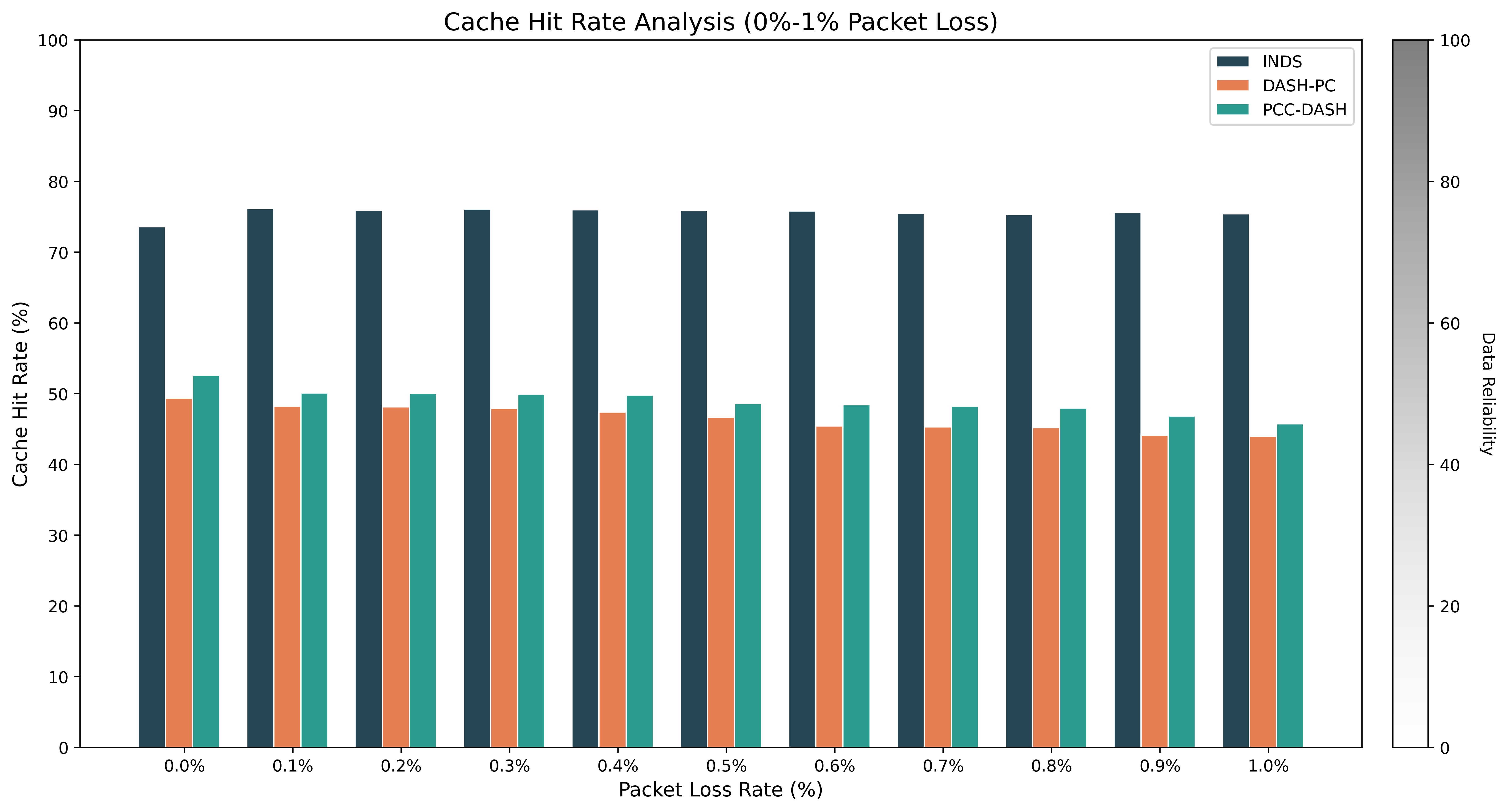}
    \caption{Cache hit rate comparison under varying packet loss rates (0\% to 1\%) for INDS, DASH-PC, and PCC-DASH.}
    \label{fig:cache_hit_rate}
\end{figure}

Figure~\ref{fig:cache_hit_rate} shows that INDS achieves a cache hit rate of 73\% to 76\% across all packet loss conditions, outperforming DASH-PC and PCC-DASH at around 50\%. This demonstrates the advantage of incremental, segment-based caching in lossy networks.

\subsection{Total Delay Evaluation}
\label{sec:delay-evaluation}
We evaluate delays in the 20–50 ms range because prior work \cite{10.1145/3523230.3523233} has shown that viewport-adaptive streaming quality degrades noticeably once latency exceeds 45 ms. 
This system-level perspective aligns with our focus on interactive volumetric streaming scenarios. 

\begin{figure*}[ht]
    \centering
    \begin{subfigure}[b]{0.32\textwidth}
        \includegraphics[width=\linewidth]{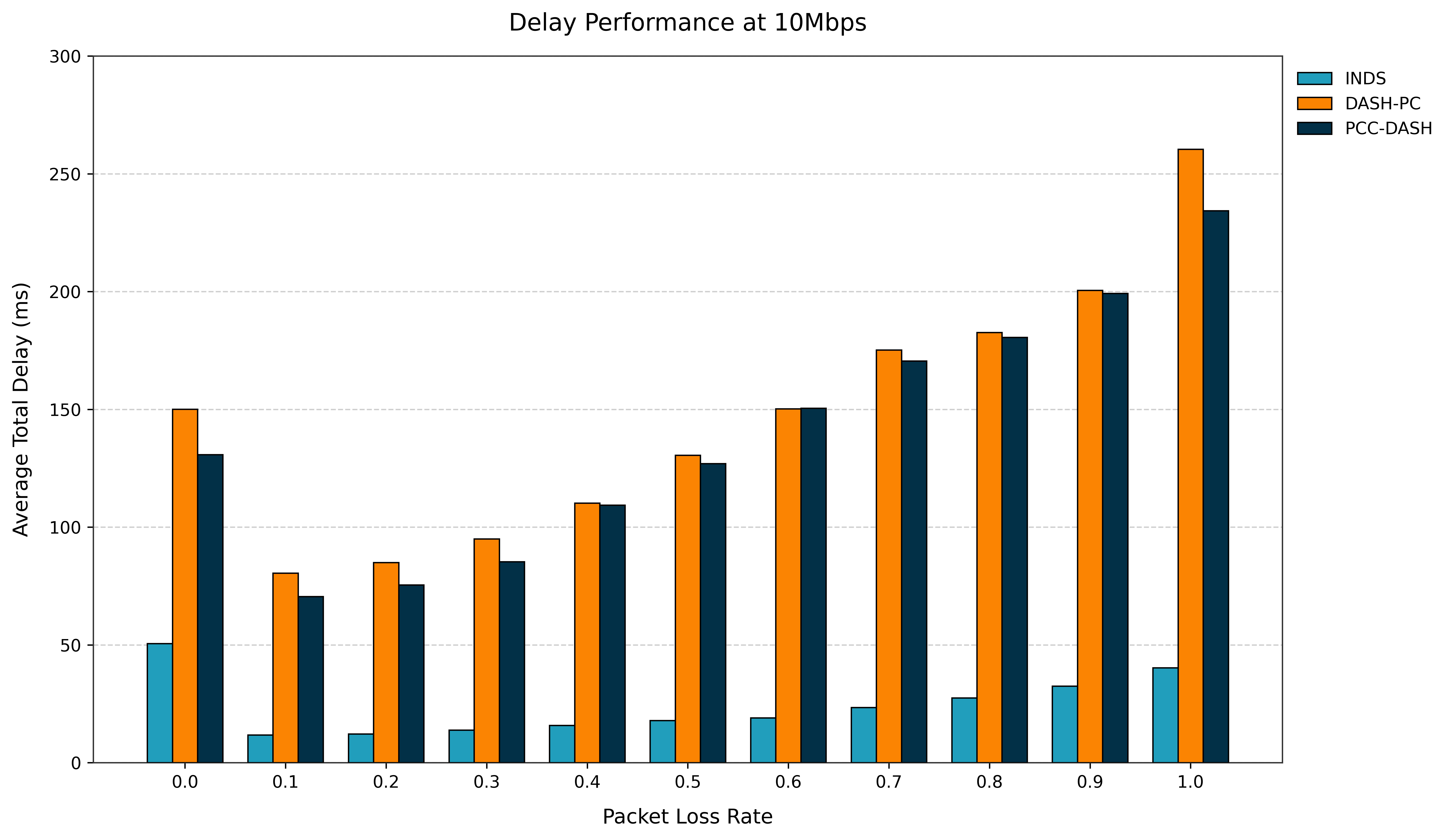}
        \caption{10 Mbps}
        \label{fig:delay_10}
    \end{subfigure}
    \hfill
    \begin{subfigure}[b]{0.32\textwidth}
        \includegraphics[width=\linewidth]{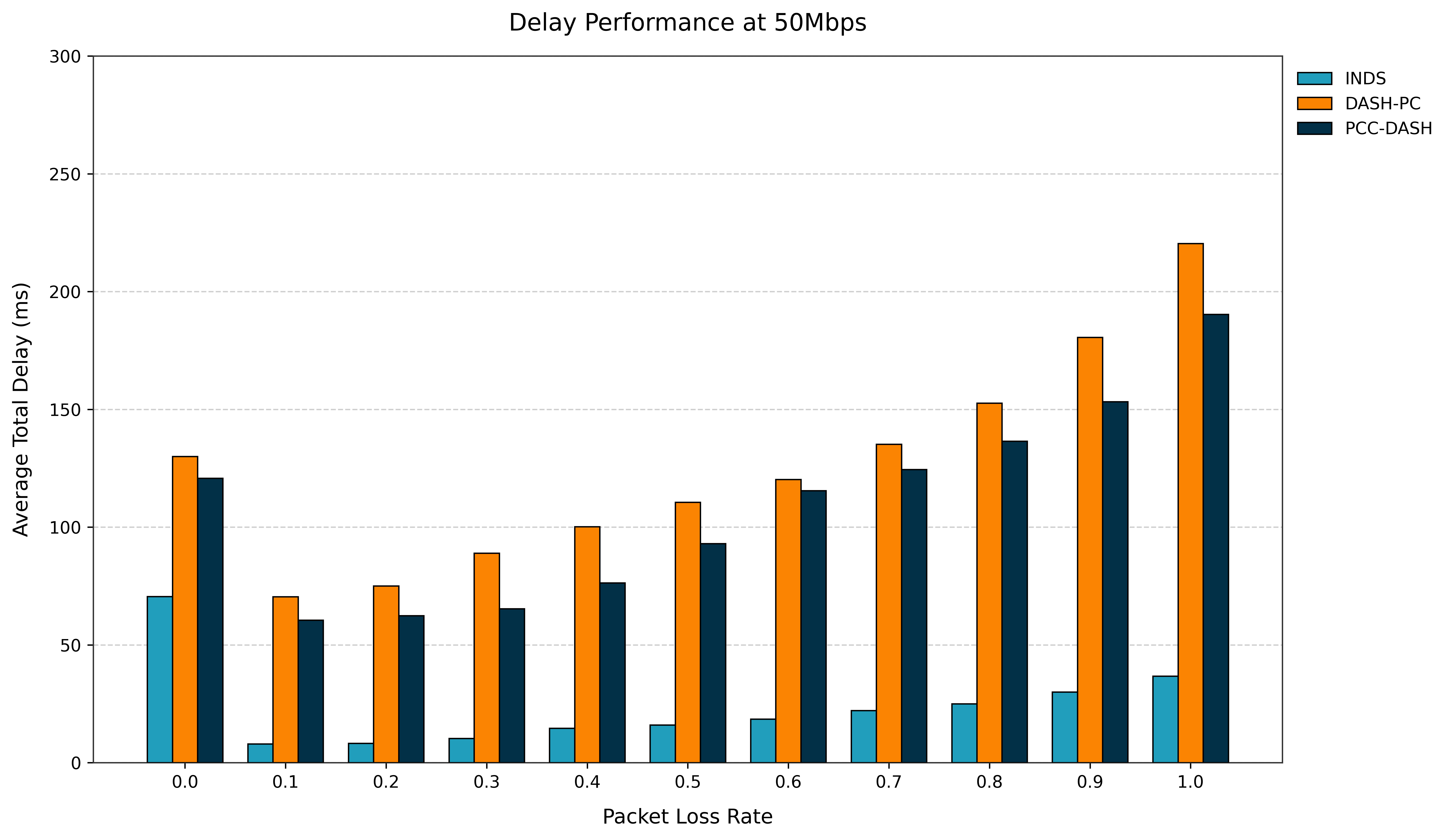}
        \caption{50 Mbps}
        \label{fig:delay_50}
    \end{subfigure}
    \hfill
    \begin{subfigure}[b]{0.32\textwidth}
        \includegraphics[width=\linewidth]{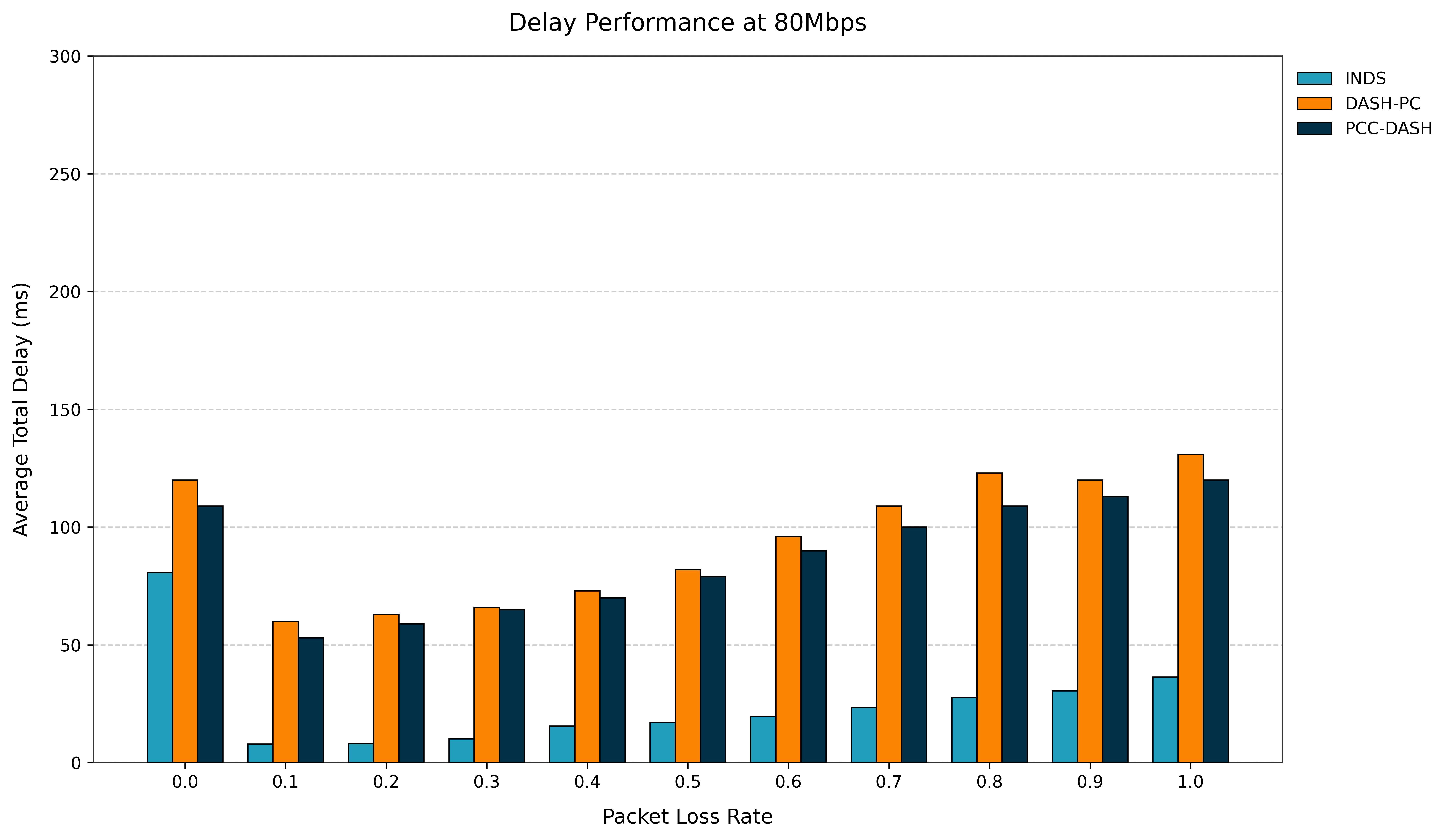}
        \caption{80 Mbps}
        \label{fig:delay_80}
    \end{subfigure}
    \caption{Total delay comparison under varying packet loss rates for (a) 10 Mbps, (b) 50 Mbps, and (c) 80 Mbps.}
    \label{fig:latency_plot}
\end{figure*}

Based on this threshold, we assess the responsiveness and robustness of INDS under diverse network conditions by measuring its total end-to-end delay at three representative downstream bandwidths: 10 Mbps, 50 Mbps, and 80 Mbps. Fig.~\ref{fig:latency_plot} compares the average total delay of INDS, DASH-PC, and PCC-DASH across different packet loss rates. 
The results consistently demonstrate that INDS achieves significantly lower and more stable delay under all conditions, 
particularly in high-loss scenarios.

\begin{figure*}[t]
    \centering
    \begin{subfigure}[b]{0.32\textwidth}
        \includegraphics[width=\linewidth]{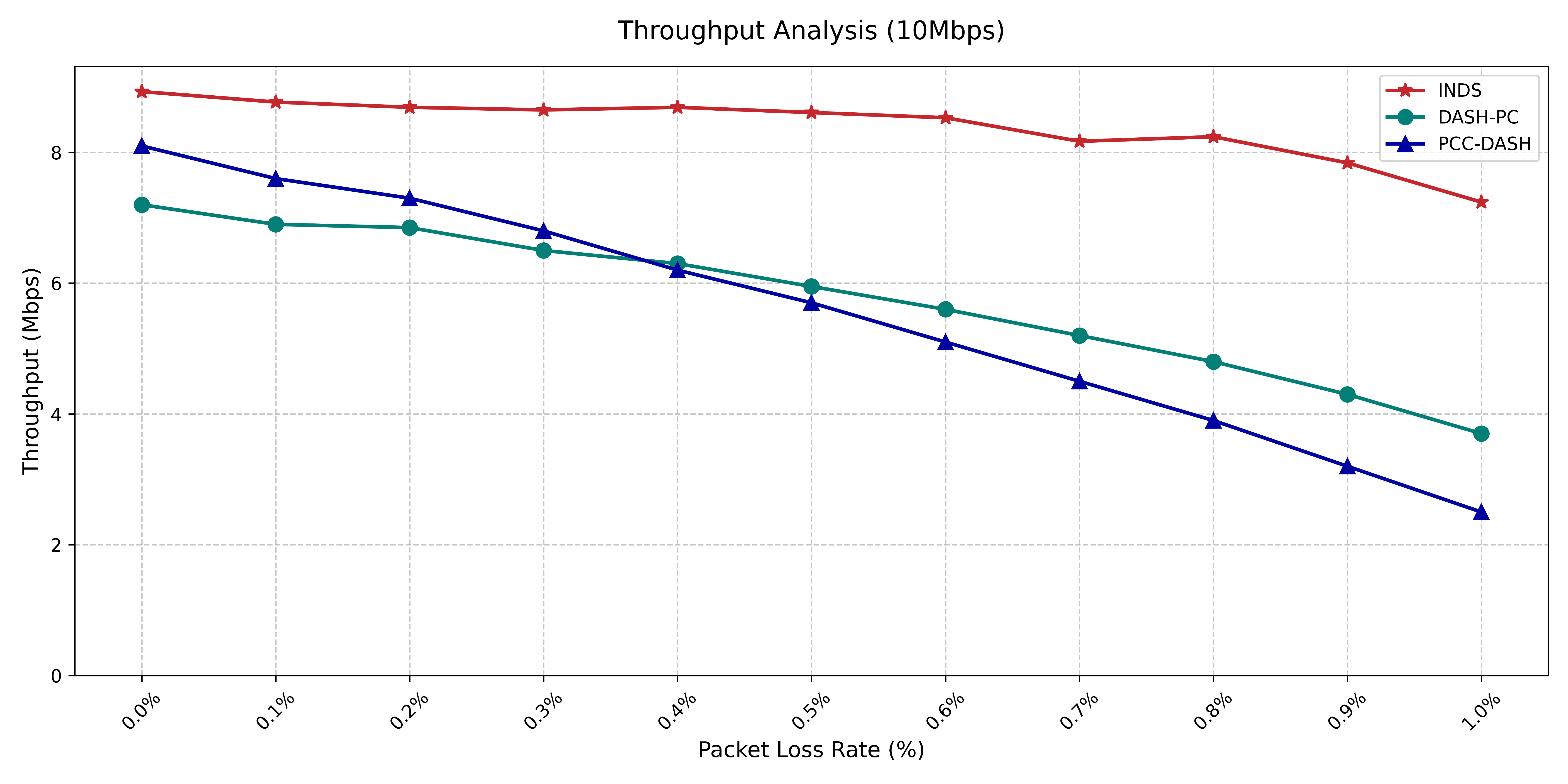}
        \caption{10 Mbps}
        \label{fig:throughput_10}
    \end{subfigure}
    \hfill
    \begin{subfigure}[b]{0.32\textwidth}
        \includegraphics[width=\linewidth]{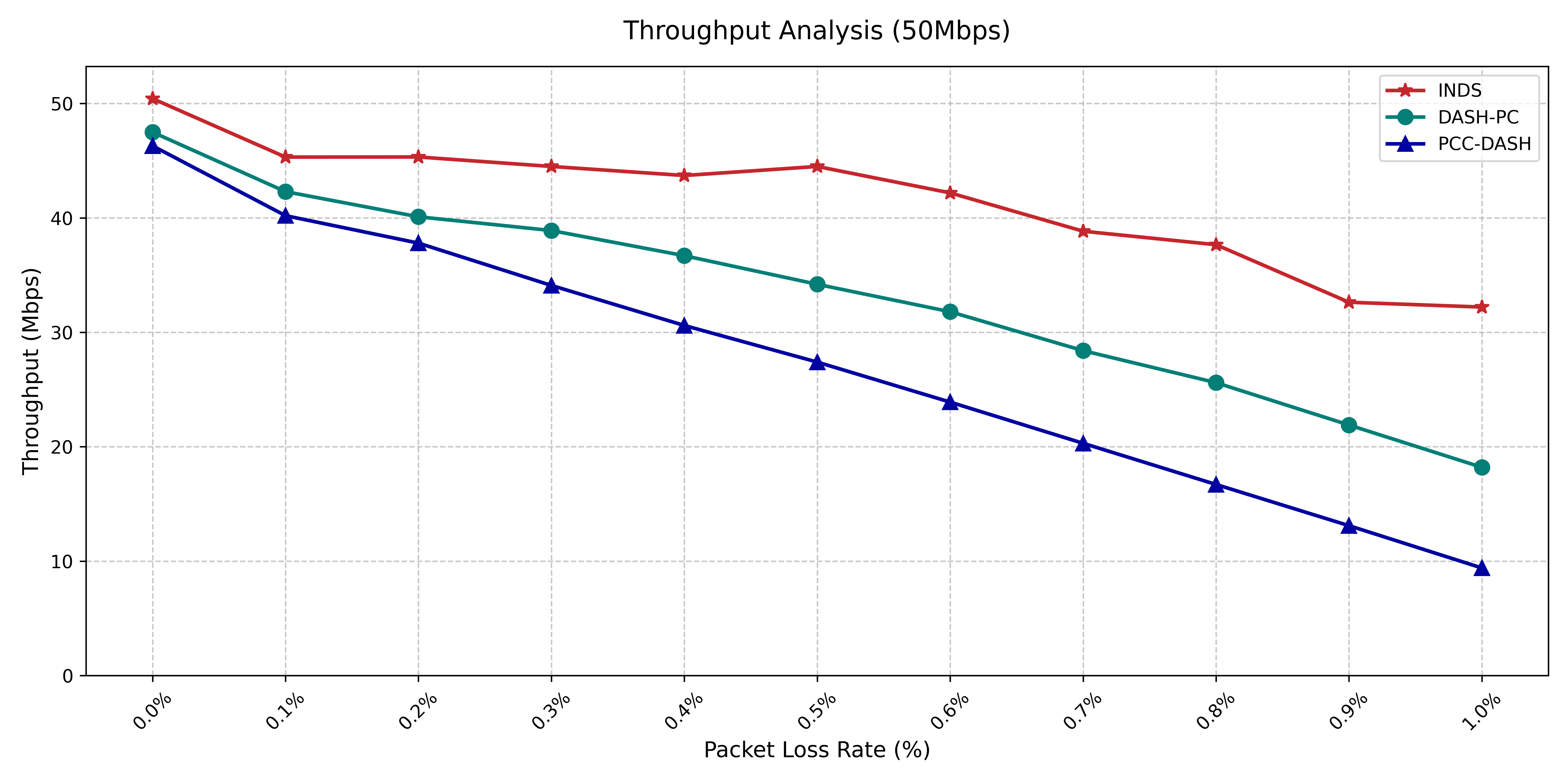}
        \caption{50 Mbps}
        \label{fig:throughput_50}
    \end{subfigure}
    \hfill
    \begin{subfigure}[b]{0.32\textwidth}
        \includegraphics[width=\linewidth]{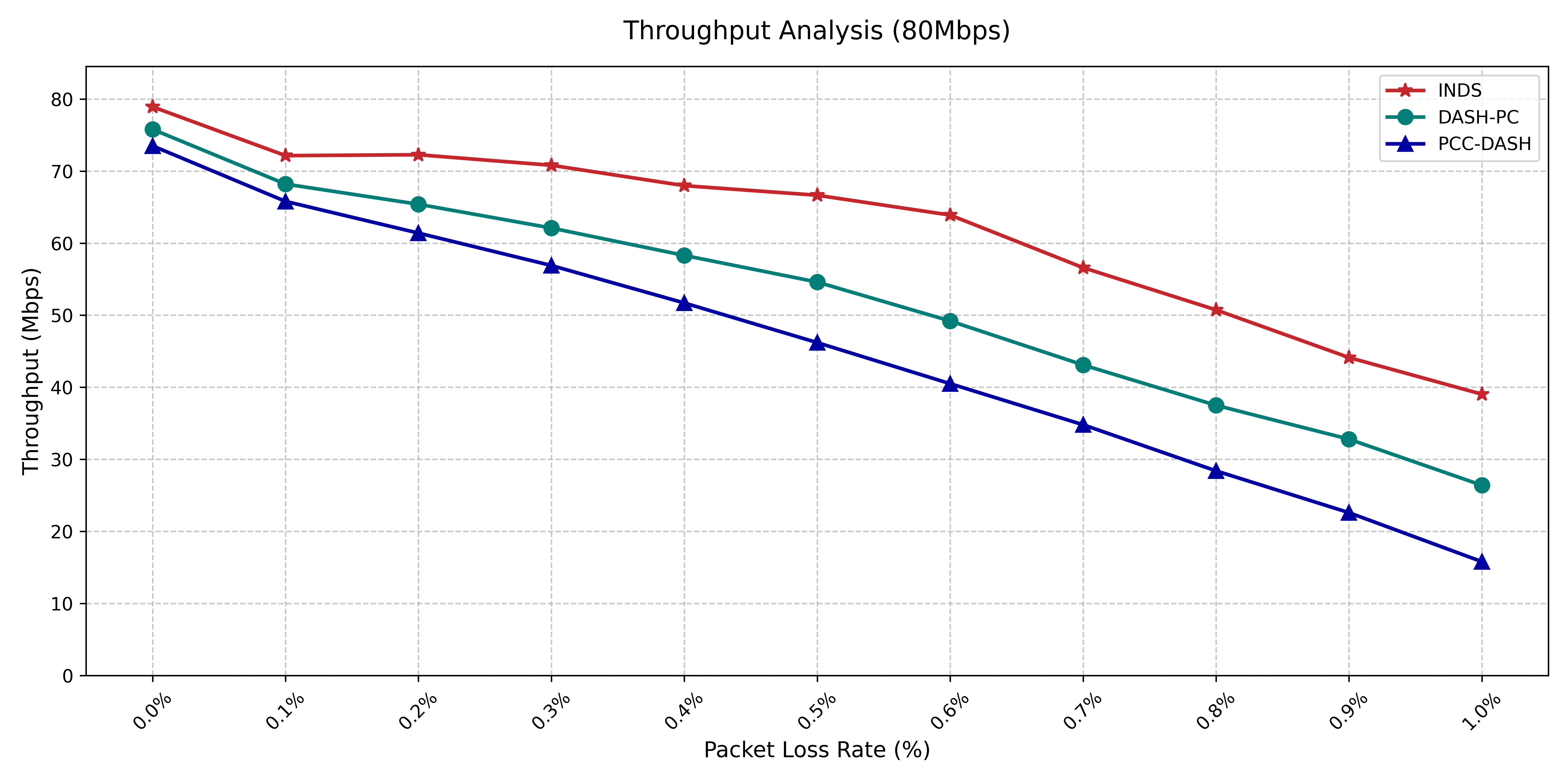}
        \caption{80 Mbps}
        \label{fig:throughput_80}
    \end{subfigure}
    \caption{Throughput performance with varying packet loss rates for (a) 10 Mbps, (b) 50 Mbps, and (c) 80 Mbps. }
    \label{fig:throughput-inds}
\end{figure*}

\subsection{End-to-End Throughput and Resilience}
\label{sec:throughput-eval}
At 10 Mbps, INDS outperforms both TCP/IP-based baselines by a wide margin—maintaining average delay below 40 ms even under 1\% packet loss, while DASH-PC and PCC-DASH experience latency spikes exceeding 250 ms and 230 ms respectively. At 50 Mbps, although all schemes exhibit improved performance, INDS still delivers up to 75\% lower delay than the baselines under heavy loss. At 80 Mbps, where network resources are less constrained, the delay gap narrows but remains in favor of INDS, which preserves its efficiency and avoids retransmission bottlenecks. These results highlight the effectiveness of INDS's fine-grained Interest/Data exchange and in-network caching. By avoiding full-segment retransmissions and enabling loss-localized recovery, INDS mitigates head-of-line blocking and supports responsive point cloud streaming. Its performance stability across loss rates and bandwidth levels confirms its suitability for real-time, delay-sensitive immersive applications.

We evaluate the effective throughput of INDS, DASH-PC, and PCC-DASH under varying packet loss conditions. Experiments are conducted at three downstream bandwidth 10 Mbps, 50 Mbps, and 80 Mbps with packet loss rates ranging from 0\% to 1\% in 0.1\% increments. Throughput is calculated as the total volume of successfully received data over the duration of each trial, excluding retransmitted or duplicated packets.

As illustrated in Fig.~\ref{fig:throughput-inds}, INDS consistently achieves higher throughput across all configurations and remains robust under increasing loss. At 10 Mbps, the throughput of DASH-PC and PCC-DASH drops by more than 45\% when packet loss exceeds 0.6\%, while INDS maintains over 85\% of its original throughput even at 1\% loss. A similar trend is observed at 50 Mbps and 80 Mbps, where the TCP-based schemes suffer from steep declines due to retransmission congestion and head-of-line blocking.

Moreover, the throughput degradation curve of INDS is notably flatter, indicating better scalability and resilience under adverse network conditions. On average, INDS preserves over 90\% of its maximum throughput under 0.5\% packet loss, whereas DASH-PC and PCC-DASH degrade to approximately 70\% and 65\%, respectively.

\section{Conclusion and Future Work}\label{section6}

We presented INDS, a novel ICN-based framework for adaptive, real-time point cloud video streaming. Unlike traditional DASH-style systems that rely on coarse-grained segment retrieval and rigid bitrate ladders, INDS leverages the hierarchical structure of Octree encoding and the expressive naming capabilities of ICN 
to enable incremental, fine-grained data retrieval. By allowing consumers to progressively request only the needed enhancement layers and exploiting in-network caching through semantically structured names, INDS reduces transmission redundancy, improves cache reuse, and maintains low latency, even in lossy networks. Our evaluation demonstrates that INDS significantly improves throughput and playback robustness compared to state-of-the-art TCP/IP-based solutions, especially under constrained conditions.

INDS is fully deployable as an overlay, without requiring changes to the core network, and aligns with ongoing developments in Media-over-QUIC (MoQ). This positions it as a practical and forward-compatible solution for emerging volumetric media systems.


Looking ahead, we plan to explore finer-grained segmentation within each enhancement layer to improve caching granularity, integrate ICN’s multipath forwarding for congestion-aware delivery, and investigate multicast extensions to better support dense user scenarios. These directions could further expand INDS’s applicability, including potential extensions to wireless and heterogeneous network environments.

In summary, INDS establishes a strong ICN-native foundation for scalable, resilient, and future-proof immersive media streaming.

\begin{acks}
This work has partly been supported by the Guangdong provincial project 2023QN10X048, Guangzhou Municipal Key Laboratory on Future Networked Systems (2024A03J0623), the Guangdong Provincial Key Lab of Integrated Communication, Sensing and Computation for Ubiquitous Internet of Things (No.2023B1212010007), the Guangzhou Municipal Science and Technology Project 2023A03J0011, the Guangdong provincial project 2023ZT10X009, and the Natural Science Foundation of China (U23A20339).
\end{acks}

\clearpage
\bibliographystyle{ACM-Reference-Format}
\balance
\bibliography{reference}

\end{document}